\documentclass[useAMS,usenatbib]{mn2e}
\usepackage[dvips]{graphicx}
\usepackage{epsfig,amssymb,amsmath,color}
\newcommand{\beq}{\begin{equation}}
\newcommand{\eeq}{\end{equation}}
\newcommand{\beqn}{\begin{eqnarray}}
\newcommand{\eeqn}{\end{eqnarray}}

\long\def\symbolfootnote[#1]#2{\begingroup%
\def\thefootnote{\fnsymbol{footnote}}\footnote[#1]{#2}\endgroup}

\usepackage{pstricks,pst-text}

\usepackage[utf8]{inputenc}
\usepackage[OT2,T1]{fontenc}
\DeclareSymbolFont{cyrletters}{OT2}{wncyr}{m}{n}
\DeclareMathSymbol{\sha}{\mathalpha}{cyrletters}{"58}

\title[Adaptive Weighting in Radio Interferometric Imaging]{Adaptive Weighting in Radio Interferometric Imaging}
\author[Yatawatta]{S. Yatawatta\\
ASTRON, Postbus 2, 7990 AA Dwingeloo, the Netherlands}
\begin{document}
\date{\today}
\pagerange{\pageref{firstpage}--\pageref{lastpage}} \pubyear{2014}
\maketitle
\label{firstpage}
%

\begin{abstract}
Radio interferometers observe the Fourier space of the sky, at locations determined by the array geometry. Before a real space image is constructed by a Fourier transform, the data is weighted to improve the quality of reconstruction. Two criteria for calculation of weights are maximizing sensitivity and minimizing point spread function (PSF) sidelobe levels. In this paper, we propose a novel weighting scheme suitable for ultra deep imaging experiments. The proposed weighting scheme is used to maximize sensitivity while minimizing PSF sidelobe variation across frequency and multiple epochs. We give simulation results that show the superiority of the proposed scheme compared with commonly used weighting schemes in achieving these objectives.  
\end{abstract}

\section{Introduction}
There are several deep imaging experiments using radio interferometers that are underway or are being planned, especially at low frequencies. A case in point for scientific motivation for such experiments is the statistical detection and imaging of the Epoch of Reionization \citep{SZ2012}. In such experiments, several hundred hours of data of the same field in the sky is collected over a wide frequency range and at different epochs. The location of sampling points of the interferometer (also called the uv coverage) is determined by the configuration of the array. These locations scale with frequency and also change over time due to flagged data (radio frequency interference) and due to precession and nutation of the Earth. Therefore, the uv coverage is not invariable and can lead to complications in extraction of science \citep{Hazelton}. Furthermore, the uv coverage is not regular and in order to use the fast Fourier transform (FFT), gridding of the data \citep{Schwab78, WProj} onto a regular grid is required.

It can be argued that by designing an array (or the location of the stations) such that the uv coverage is completely filled and regular, the aforementioned problems can be mitigated. However, even in this case, the density of the sampling points in the Fourier space (uv plane) might not be uniform. Therefore, a practical solution is to select appropriate data weights (also called imaging weights or density compensation factors) to get the desired data sampling density. An overview of existing weighting schemes used in radio interferometry can be found in \cite{Briggs} and \cite{Boone}. Commonly used 'natural' weights make the weights inversely proportional to the noise variance at each sampling point. Given data with uniformly distributed noise, this is equivalent to making the weights equal to unity. On the other hand 'uniform' weights make the weights inversely proportional to the number of data points within a gridded cell. It is well noted \citep{Briggs,Boone} that natural weights yield the maximum sensitivity while uniform weights yield the minimum PSF sidelobe levels. An intermediate 'robust' weighting scheme was proposed by \cite{Briggs} that can trade off sensitivity with sidelobe level or vice versa. Recently, \cite{Boone} proposed a similar scheme by parametrizing error between desired (ideal) PSF and actual PSF and selecting parameters to suit the need. It should be noted that the method proposed by \cite{Boone} can be described as gridding data with uniform weights and applying additional 'tapering' to the gridded data to get the desired PSF. Considering all such weighting schemes, we note that: (i) The weight calculation is done in one pass over the data, i.e. there is no iterative update of weights. (ii) For uniform and robust weights the data points that fall within a gridded cell are given almost the same weights, i.e. the variation of the weights over spatial scales larger than the field of view (FOV) (the field of view is inversely proportional to the size of gridded cells) is minimal. 

At this point, we like to emphasize the close similarity between radio interferometric imaging and magnetic resonance imaging. In the latter, there is a wide variety of iterative weighting techniques that are in use \citep{PM,Wajer99,Johnson,Samsonov}. Hence it is straight forward to adopt these techniques to radio interferometric imaging. In this paper, we use the method proposed by \cite{PM} to find an improved weighting scheme for radio interferometric imaging. Our primary objective in this paper is to find a weighting scheme that maximizes the sensitivity whilst minimizing the PSF variation over frequency or over different epochs. As shown in simulations, the proposed 'adaptive' weighting scheme is not inherently liked to the geometry of the gridded cells. Therefore, the variation of weights over the uv plane is much smoother. This will benefit deep imaging science  \citep{SZ2012,Hazelton} and overcome some problems therein. Moreover, the proposed scheme enables us to fine-tune the PSF with more freedom. Considering that radio interferometric images are a convolution of the true sky with the PSF, it is possible to fine-tune the PSF to enhance certain aspects of the image (similar to matched { filter techniques} \citep{Turin60}). Despite these advantages of the proposed weighting scheme, we note that it is computationally more expensive than existing approaches, but this can be overcome by using parallelized algorithms.

The rest of the paper is organized as follows: In section \ref{sec:gridding} we give an overview of imaging using convolutional gridding and existing weighting schemes. In section \ref{sec:adaptive} we describe the adaptive weighting scheme proposed in this paper and give simulation results in section \ref{sec:simul} to illustrate its benefits. Finally, we draw our conclusions in section \ref{sec:conclusions}.

\section{Convolutional Gridding}\label{sec:gridding}
In this section we give a brief overview of imaging based on convolutional gridding of the data  \citep{Brouw75,Schwab80}. We also give a brief overview of commonly used weighting schemes in radio interferometry. For data with three dimensional sampling, the following description equally holds except that the data is projected onto the $w=0$ plane \citep{WProj}. We denote the coordinates on the uv plane as $(u,v)$ and on the image plane as $(l,m)$. Let the sky signal be $s(l,m)$ in real space and its corresponding signal on the Fourier plane be $S(u,v)$. Note that by conjugate symmetry, we also treat $S(-u,-v)$ (conjugated) as an independent data point. The data weights are given by $W(u,v)$ and $w(l,m)$ is its image plane equivalent. We denote the convolution kernel by $C(u,v)$ and its Fourier transform by $c(l,m)$. We use prolate spheroidal wave functions \citep{PSWF} as our convolutional kernel throughout the paper.

The locations where $S(u,v)$ is available are discrete, sparse and irregular due to the array configuration. By weighting and convolving with the convolutional kernel, we get its continuous representation as $(S(u,v)W(u,v)) \otimes C(u,v)$. This is sampled onto a regular grid of cells in order to take the FFT. Assuming the gridded cell size to be $B$, the gridded data is given as
\beq \label{sgrid}
\tilde{S}(u,v)=\left( (S(u,v)W(u,v)) \otimes C(u,v) \right)  \sha(u/B,v/B) 
\eeq
where $\sha(u/B,v/B)$ is the Dirac comb with uv cell size $B$. After taking the FFT of the gridded data we get
\beq
\tilde{s}(l,m)= \left( (s(l,m)\otimes w(l,m)) c(l,m) \right) \otimes \sha(Bl,Bm)
\eeq
where $\sha(Bl,Bm)$ is the Fourier transform of $\sha(u/B,v/B)$ (scale factor ignored). If $1/B$ is large enough, assuming no aliasing, and after apodization correction, the final image is $s(l,m)\otimes w(l,m)$. Therefore, the PSF is the Fourier transform of the weights, or $w(l,m)$.

In commonly used 'natural' weighting, the weight of the $i$-th sample in the uv plane is chosen as
\beq \label{Wnatural}
W_n(u_i,v_i) \propto \frac{1}{\sigma_{u_i,v_i}^2}
\eeq
where $\sigma_{u_i,v_i}^2$ is the noise variance of $S(u_i,v_i)$, which is generally assumed to be equal to $1$.
On the other hand, in 'uniform' weighting the weight of the $i$-th sample is chosen to be inversely proportional to the number of data points that fall within the grid cell that the $i$-th sample belongs to. To elaborate this, let cell $(p,q)$ be the grid cell within which the $i$-th sample is included. That is to say that $pB-B/2 \le u_i < pB+B/2$ and $qB-B/2 \le v_i < qB+B/2$ for integer values of $p$ and $q$. If the total number of data points that are included in grid cell $(p,q)$ is $T_{p,q}$, the 'uniform' weight of the $i$-th sample is
\beq \label{Wuniform}
W_u(u_i,v_i) \propto \frac{1/\sigma_{u_i,v_i}^2}{T_{p,q}}.
\eeq

The method proposed by \cite{Briggs} uses an additional user defined parameter to vary the weights between $W_n(u_i,v_i)$ and $W_u(u_i,v_i)$. Moreover, the method of \cite{Boone} can be considered equivalent to first selecting $W_u(u_i,v_i)$ as the weights, gridding the data, and then applying additional weights to the gridded data cells (or tapering).

\section{Adaptive Weighting}\label{sec:adaptive}
In this section, we describe the proposed 'adaptive' weighting scheme, which is directly derived from \cite{PM}. The motivation for \cite{PM} to derive an iterative weighting scheme is as follows. Assume we need to closely approximate a desired function $g(l,m)$ as the PSF in image space, 
\beq \label{approx}
w(l,m) c(l,m) \approx g(l,m).
\eeq
We assume that $g(l,m)$ is bandlimited and has finite support and since the convolutional kernel $c(l,m)$ also has similar properties we prefer (\ref{approx}) over writing it as $w(l,m)\approx g(l,m)$. We convolve both sides of (\ref{approx}) by $w(l,m)$ to get
\beq \label{wconv}
w(l,m) \otimes\left( w(l,m) c(l,m) \right) \approx w(l,m) \otimes g(l,m).
\eeq
Taking Fourier transform of both sides of (\ref{wconv}), we get
\beq \label{Wconv}
W(u,v) \left(  W(u,v) \otimes C(u,v) \right) \approx W(u,v) G(u,v)
\eeq
where $G(u,v)$ is the Fourier transform of $g(l,m)$, and assuming $\left( W(u,v) \otimes C(u,v) \right)$ is finite, we divide (\ref{Wconv}) by this term to get 
\beq \label{Wnew}
W(u,v) \approx \frac{W(u,v) G(u,v)}{\left( W(u,v) \otimes C(u,v) \right) }.
\eeq
We can use (\ref{Wnew}) to find $W(u,v)$ such that, when applied to the data, we can closely approximate $g(l,m)$. Iteratively, this can be expressed as
\beq \label{Witer}
W^{j+1}(u_i,v_i) \leftarrow \frac{W^{j}(u_i,v_i) G(u_i,v_i)}{\left(  W^{j}(u,v)  \otimes C(u,v) \right)|_{u_i,v_i} }
\eeq
where $W^{j}(u_i,v_i)$ is the weight of the $i$-th sample at the $j$-th iteration. We calculate the right hand side of (\ref{Witer}) using the weights of the $j$-th iteration and update the weights for the $j+1$-th iteration. In other words, given $G(u,v)$, and given (\ref{Witer}) converges, we can iteratively find $W(u,v)$ to satisfy (\ref{approx}). For instance, when $g(l,m)=\delta(l,m)$ which is the Dirac delta function, we use $G(u,v)=1$ and we get 'uniform' weights for $W(u,v)$. For convergence of (\ref{Witer}), we need both $G(u,v)$ and $C(u,v)$ to be positive. The weights $W(u,v)$ are always assumed to be positive and are initialized to be $1$. There are improvements made to (\ref{Witer}) to increase the speed of convergence \citep{Samsonov,Johnson,Zwart} but in this paper we restrict ourselves to (\ref{Witer}).

\begin{figure}
\begin{minipage}{0.98\linewidth}
\begin{minipage}{0.98\linewidth}
\centering
 \centerline{\epsfig{figure=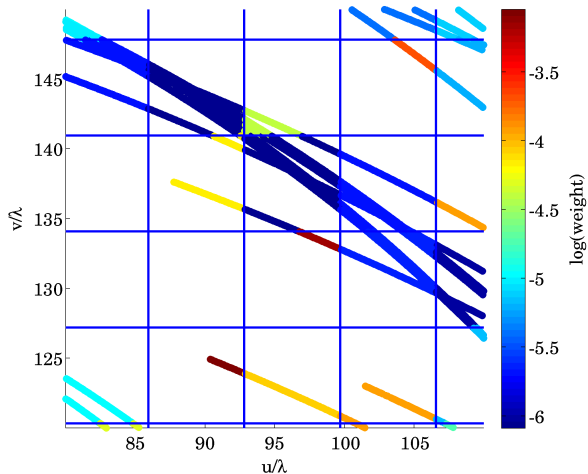,width=8.0cm}}
\vspace{0.5cm} \centerline{(a)}\smallskip
\end{minipage}\\
\begin{minipage}{0.98\linewidth}
\centering
 \centerline{\epsfig{figure=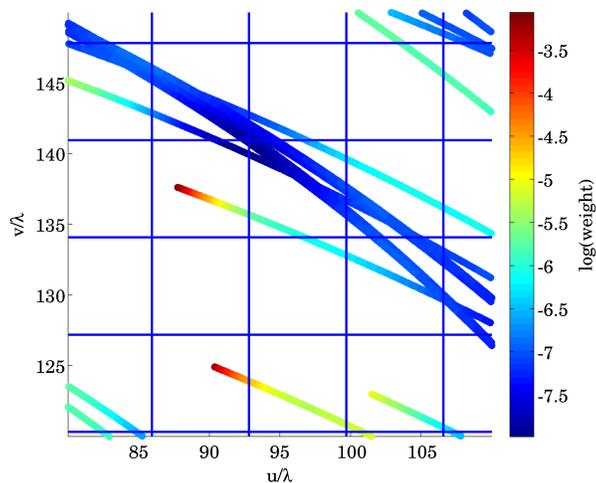,width=8.0cm}}
\vspace{0.5cm} \centerline{(b)}\smallskip
\end{minipage}
\caption{Image weights (log-scale) on an area covering $4.5\times 4.5$ grid cells. (a) Uniform weights, where points within each grid cell are assigned equal weights. (b) Adaptive weights obtained by (\ref{Witer}), after 10 iterations. In (b), within a grid cell, the weights are smoothly varying while there is no such variation in (a).\label{Ex}}
\end{minipage}
\end{figure}

\begin{figure}
\begin{minipage}{0.98\linewidth}
\centering
 \centerline{\epsfig{figure=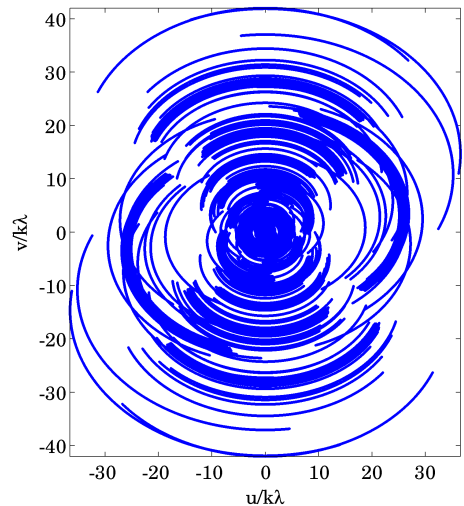,width=8.0cm}}
\caption{Simulated uv coverage, which is not completely filled.\label{uvcov}}
\end{minipage}
\end{figure}

\begin{figure}
\begin{minipage}{0.98\linewidth}
\begin{center}
\begin{minipage}{0.98\linewidth}
\centering
 \centerline{\epsfig{figure=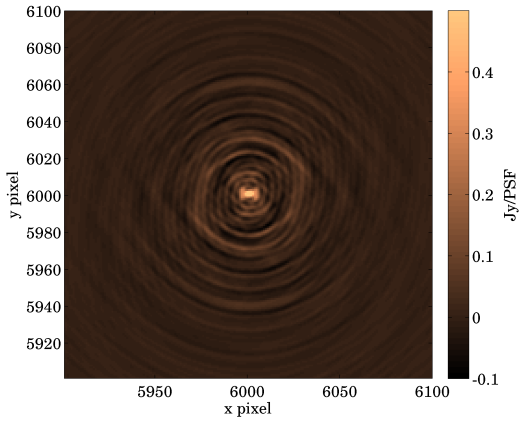,width=8.0cm}}
\vspace{0.5cm} \centerline{(a)}\smallskip
\end{minipage}
\begin{minipage}{0.98\linewidth}
\centering
 \centerline{\epsfig{figure=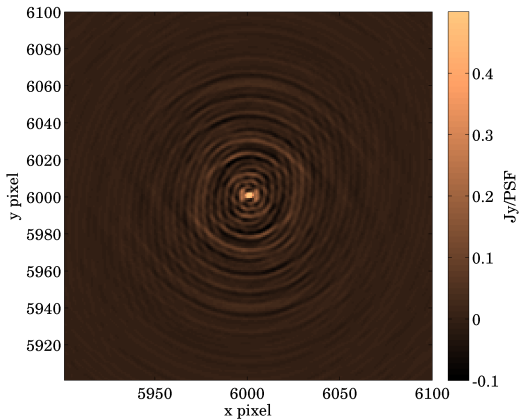,width=8.0cm}}
\vspace{0.5cm} \centerline{(b)}\smallskip
\end{minipage}
\end{center}
\caption{{ PSFs obtained by (a) Uniform weights and by (b) Adaptive weights with $G(u,v)=1$. The pixel size is $2^{\prime\prime}$. The peak difference between the two is less than 3\% of the peak value of the PSF.}\label{psferror_unifpm}}
\end{minipage}
\end{figure}

In order to compare existing weighting schemes with the one proposed in this section, we give an example in Fig. \ref{Ex}. In this figure we show weights of a small area of about 4.5$\times$4.5 grid cells in the uv plane { (the full uv coverage is shown in Fig. \ref{uvcov}). The uv coverage is incomplete (i.e. not filled)}. We have shown weights calculated using traditional 'uniform' scheme and weights calculated using (\ref{Witer}) with $G(u,v)=1$ in Fig. \ref{Ex}. We see that in Fig. \ref{Ex} (a), equal weights are assigned to points within a grid cell. However, in Fig. \ref{Ex} (b), the weights vary smoothly both globally and within each gridded cell. The PSFs obtained by both weighting schemes { are shown in Fig. \ref{psferror_unifpm} with image pixel size of 2$^{\prime\prime}$.}

\begin{figure}
\begin{minipage}{0.98\linewidth}
\centering
 \centerline{\epsfig{figure=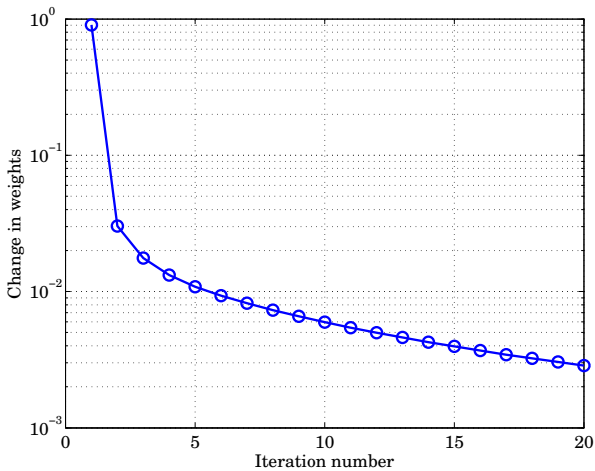,width=8.0cm}}
\caption{Normalized variation of the weights (\ref{var})  with iteration number.\label{convergence}}
\end{minipage}
\end{figure}

The convergence of the weights evaluated by (\ref{Witer}) is shown in Fig. \ref{convergence}. In this figure we have plotted the change in the total weight 
\beq \label{var}
E_j=\frac{\sqrt{\sum_i (W^{j+1}(u_i,v_i)-W^j(u_i,v_i))^2}}{\sqrt{\sum_i (W^j(u_i,v_i))^2}}
\eeq
where $j$ is the iteration number. We have only used 10 iterations for the comparison in Fig. \ref{Ex}.

At this juncture, we make several observations: First, by properly selecting $g(l,m)$, the proposed adaptive weighting scheme is able to control the weights in a more elaborate way than what can be done using conventional weights. Secondly, we see no sharp jumps in the weight distribution of adaptive weights, provided that $G(u,v)$ is smooth. These two properties are useful in deep imaging experiments \citep{SZ2012,Hazelton}. By keeping $g(l,m)$ fixed over multiple frequencies and over different epochs, we can expect to get almost the same PSF. Moreover, due to the weakness of the sought after signal, it is also important to maximize the sensitivity to certain scales in the image plane. This can be achieved by selecting $g(l,m)$ { such that the PSF acts as a} matched filer to those scales. { For instance, given the signal of interest $F(u,v)$, we can select $g(l,m)$ such that we get $W(u,v)\approx |F(u,v)|$. Moreover, this also involves fine-tuning the convolution kernel and further investigation of this topic is left as future work.}

The only drawback of the adaptive weighting scheme is its computational cost. Not only is it iterative, but there is a convolution in (\ref{Witer}). However, we note that the weight update of one data point is only dependent on adjacent data points because the convolutional kernel has finite support. We can use this to parallelize the evaluation of (\ref{Witer}) to speed up convergence. There are also improved algorithms \citep{Samsonov,Johnson,Zwart} that can be used to get faster results.

\section{Simulations}\label{sec:simul}
\begin{figure}
\begin{minipage}{0.98\linewidth}
\centering
 \centerline{\epsfig{figure=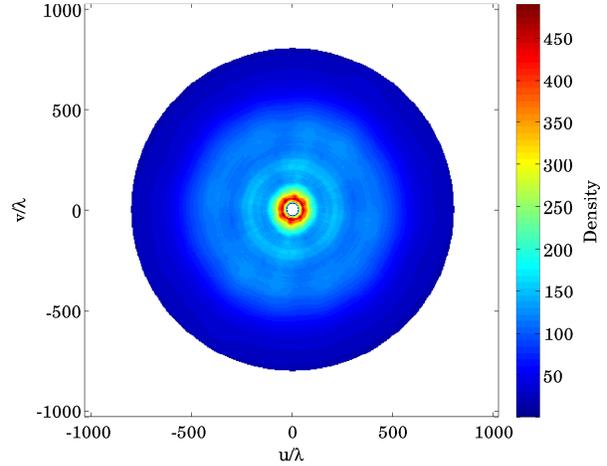,width=8.0cm}}
\caption{Density of the sampling points on the uv plane over the full observing bandwidth (115 MHz to 175 MHz).\label{density_cmap}}
\end{minipage}
\end{figure}

In this section, we give detailed simulations to measure the performance of the adaptive weighting scheme, in comparison with conventional weighting schemes. Our objective is to minimize the PSF variation over a large bandwidth (115 MHz to 175 MHz) whilst maximizing the sensitivity. We choose the array configuration to be similar to LOFAR observing a  field centered at the north celestial pole and more specific detail can be found in \cite{NCP2013}. We select baselines in the range 30$\lambda$ to 800$\lambda$ where $\lambda$ is the wavelength. The imaging parameters are set to 30$^{\prime\prime}$ pixel size with 1200$\times$1200 pixels. The density of the sampling points over the full bandwidth is shown in Fig. \ref{density_cmap}. The uv plane is 'filled' but the density is not uniform.

\begin{figure}
\begin{minipage}{0.98\linewidth}
\centering
 \centerline{\epsfig{figure=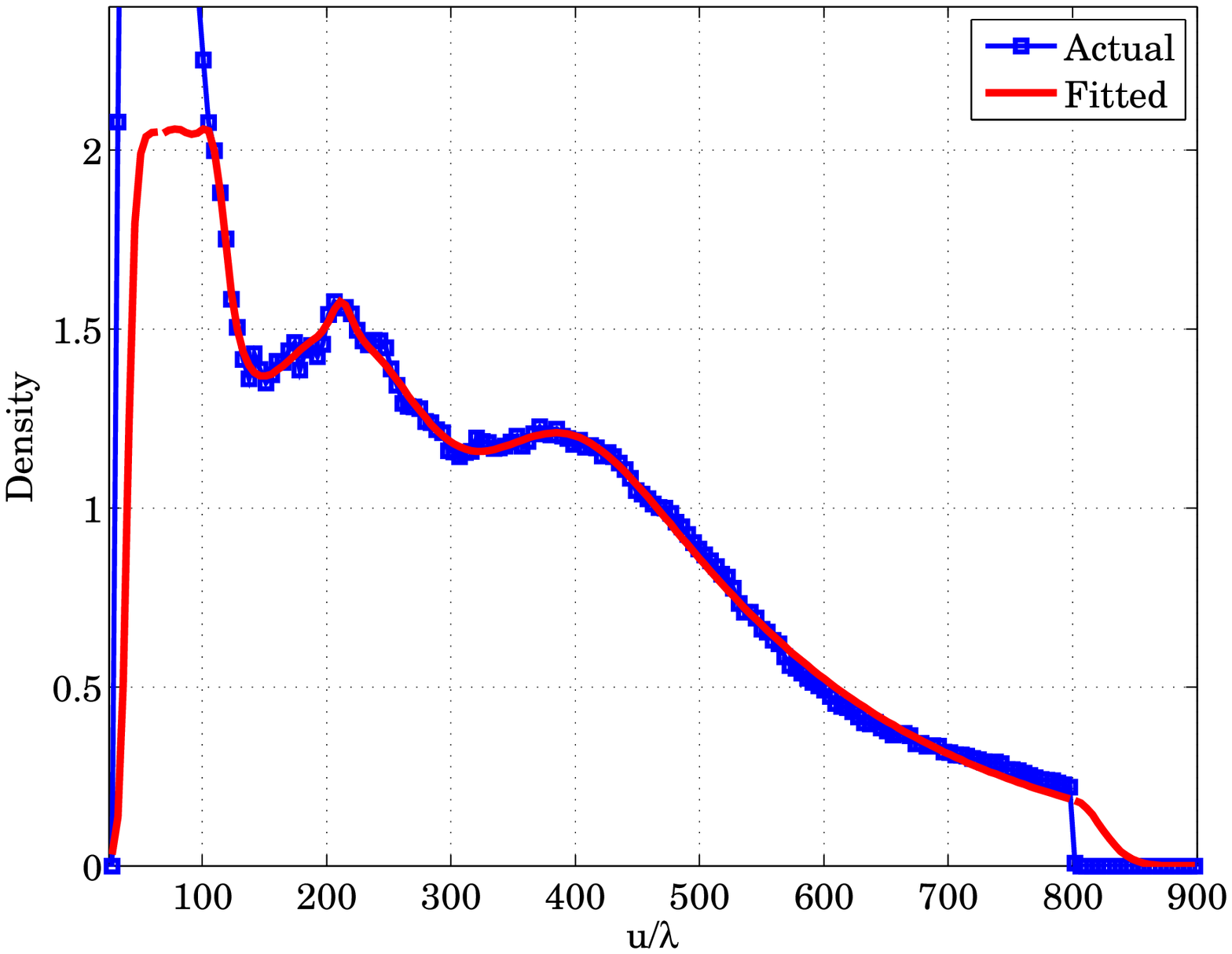,width=8.0cm}}
\caption{Radial profile of the density of sampling points and a polynomial fit for this density. The inner and outer boundaries have additional tapering polynomials.\label{density}}
\end{minipage}
\end{figure}

In order to maximize the sensitivity, we prefer to use all data samples without any down weighting. Therefore, we select the density map itself, after scaling to have mean value of about 1, as $G(u,v)$ for this simulation. We can also exploit the radial symmetry seen in Fig. \ref{density_cmap} and parametrize $G(u,v)$ as $G(\sqrt{u^2+v^2})$. A radial cut of the density map is shown in Fig. \ref{density}. We have fitted a piecewise smooth polynomial to the actual density. Note that we have added tapering polynomials to the inner and outer boundaries of the density profile.

\begin{figure}
\begin{minipage}{0.98\linewidth}
\centering
 \centerline{\epsfig{figure=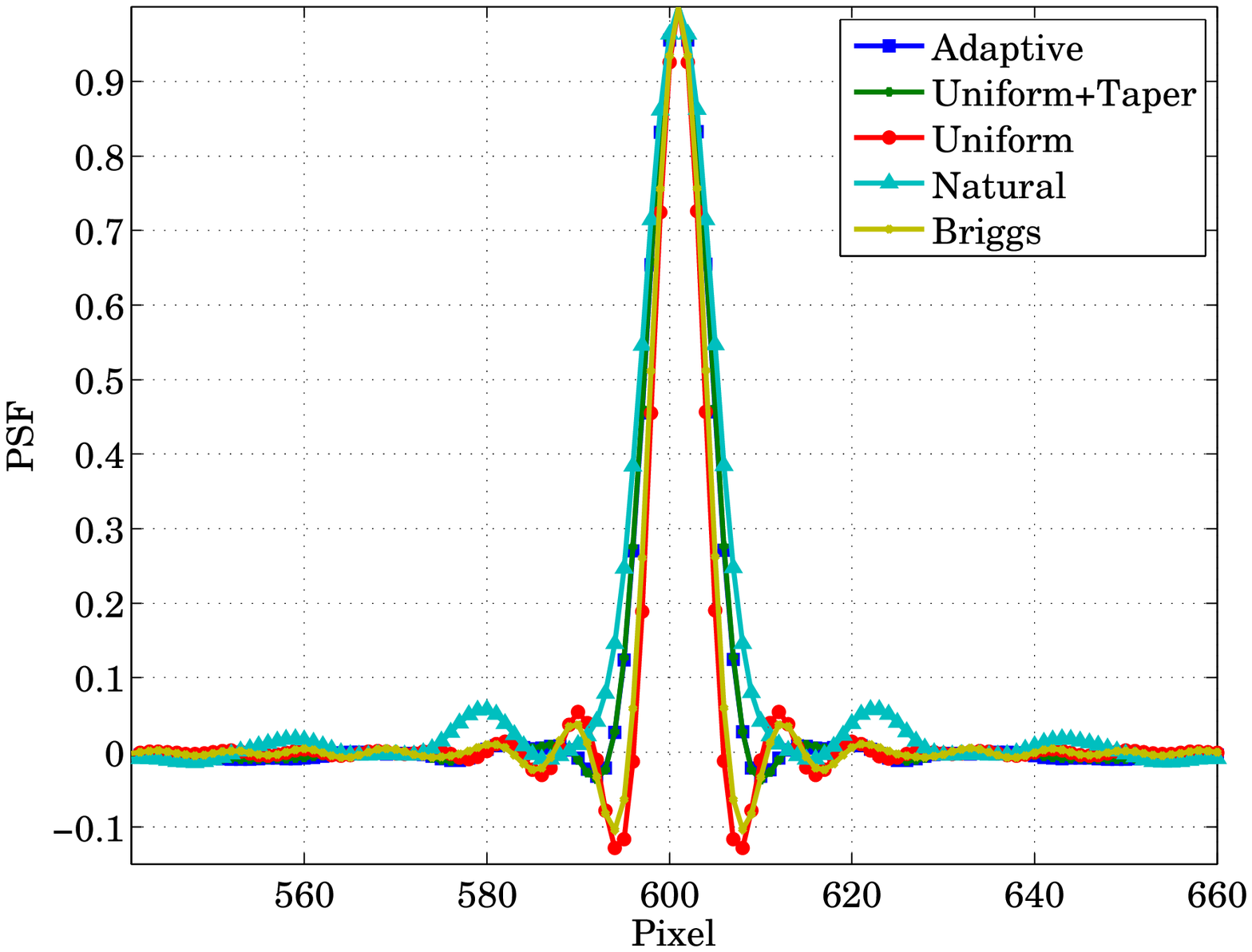,width=8.0cm}}
\caption{Cross section of the PSF obtained by the five different weighting schemes at { 115} MHz.\label{psfcross}}
\end{minipage}
\end{figure}

\begin{figure*}
\begin{minipage}{0.98\linewidth}
\begin{center}
\begin{minipage}{0.48\linewidth}
\centering
 \centerline{\epsfig{figure=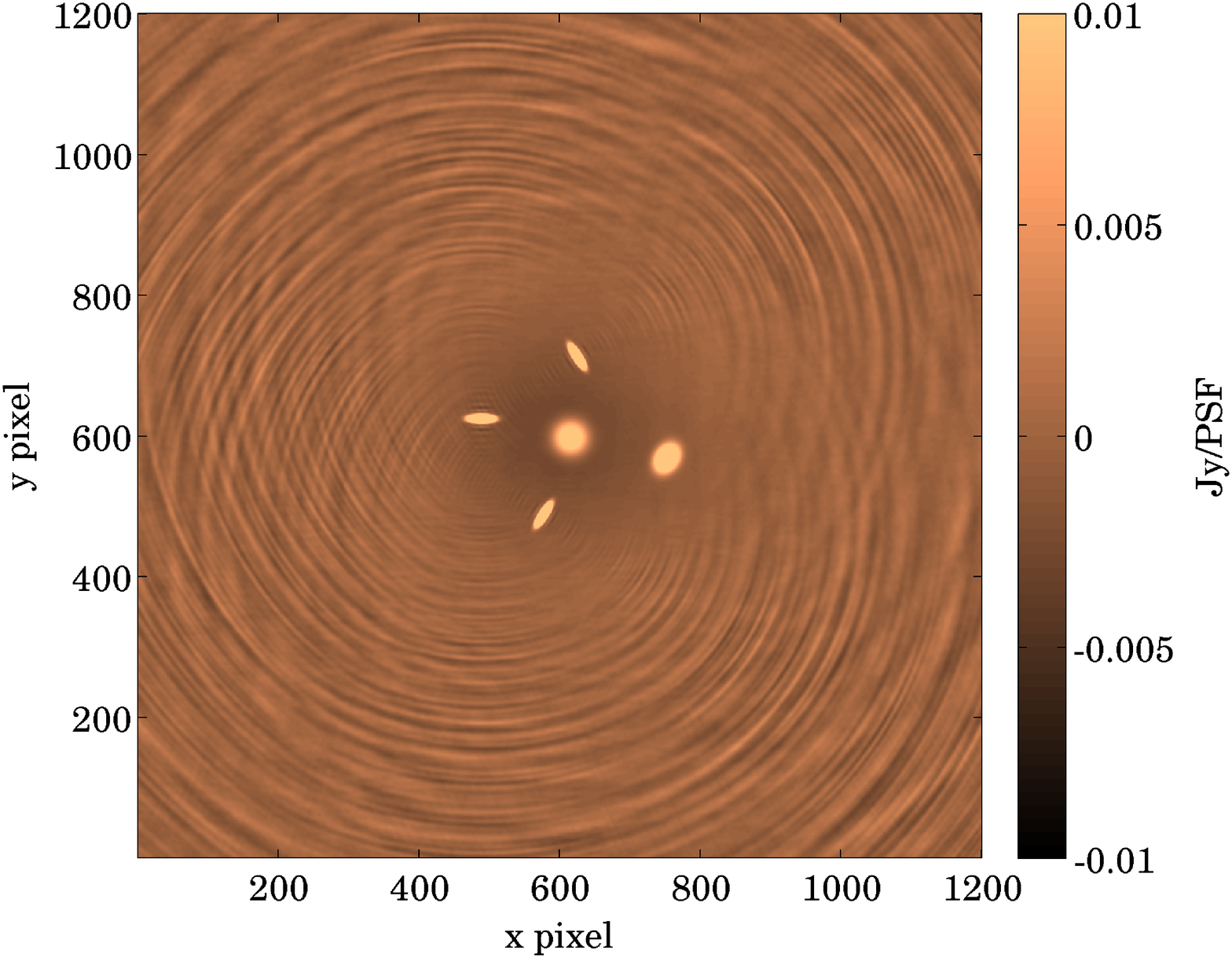,width=8.0cm}}
\vspace{0.5cm} \centerline{(a)}\smallskip
\end{minipage}
\begin{minipage}{0.48\linewidth}
\centering
 \centerline{\epsfig{figure=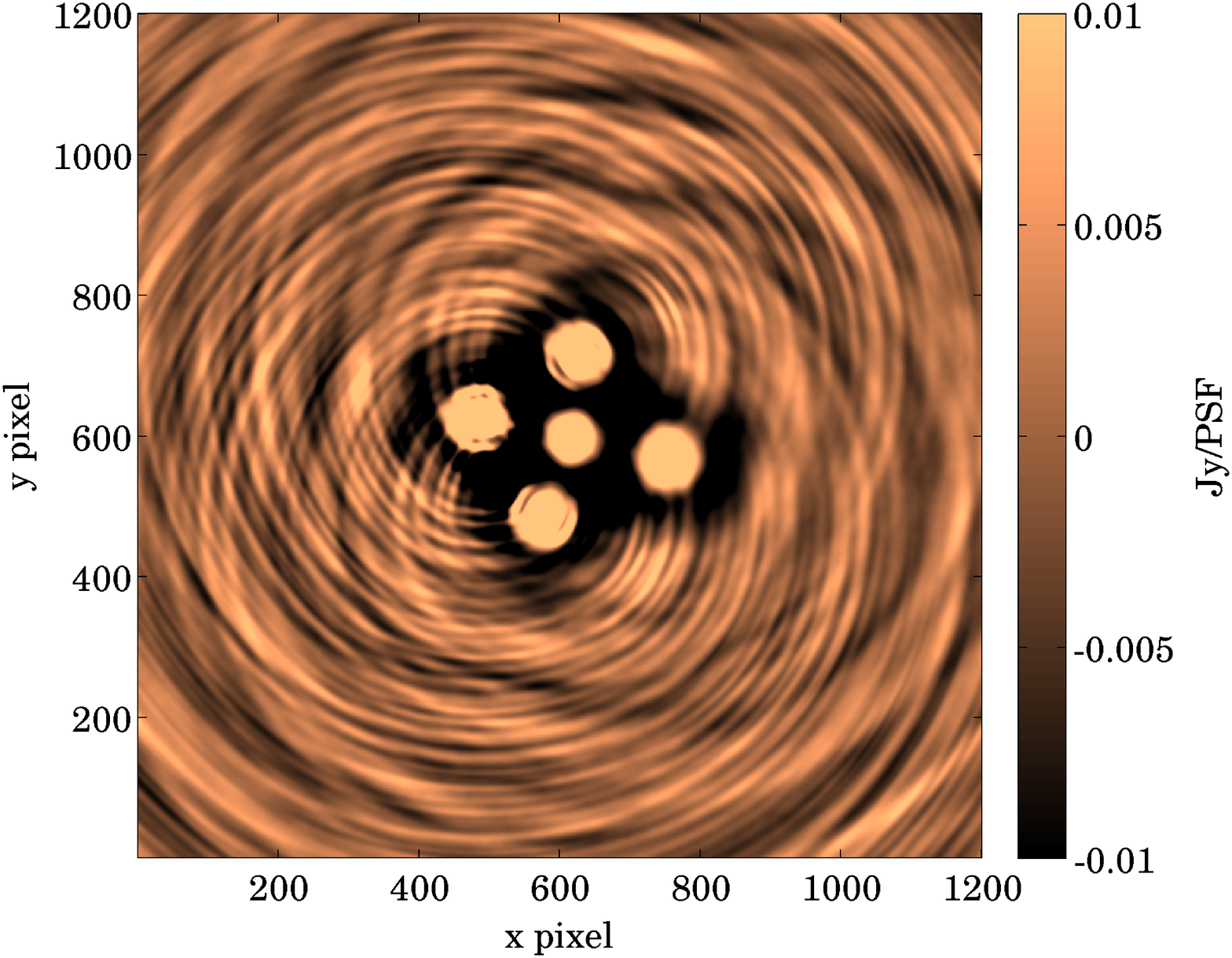,width=8.0cm}}
\vspace{0.5cm} \centerline{(b)}\smallskip
\end{minipage}
\begin{minipage}{0.48\linewidth}
\centering
 \centerline{\epsfig{figure=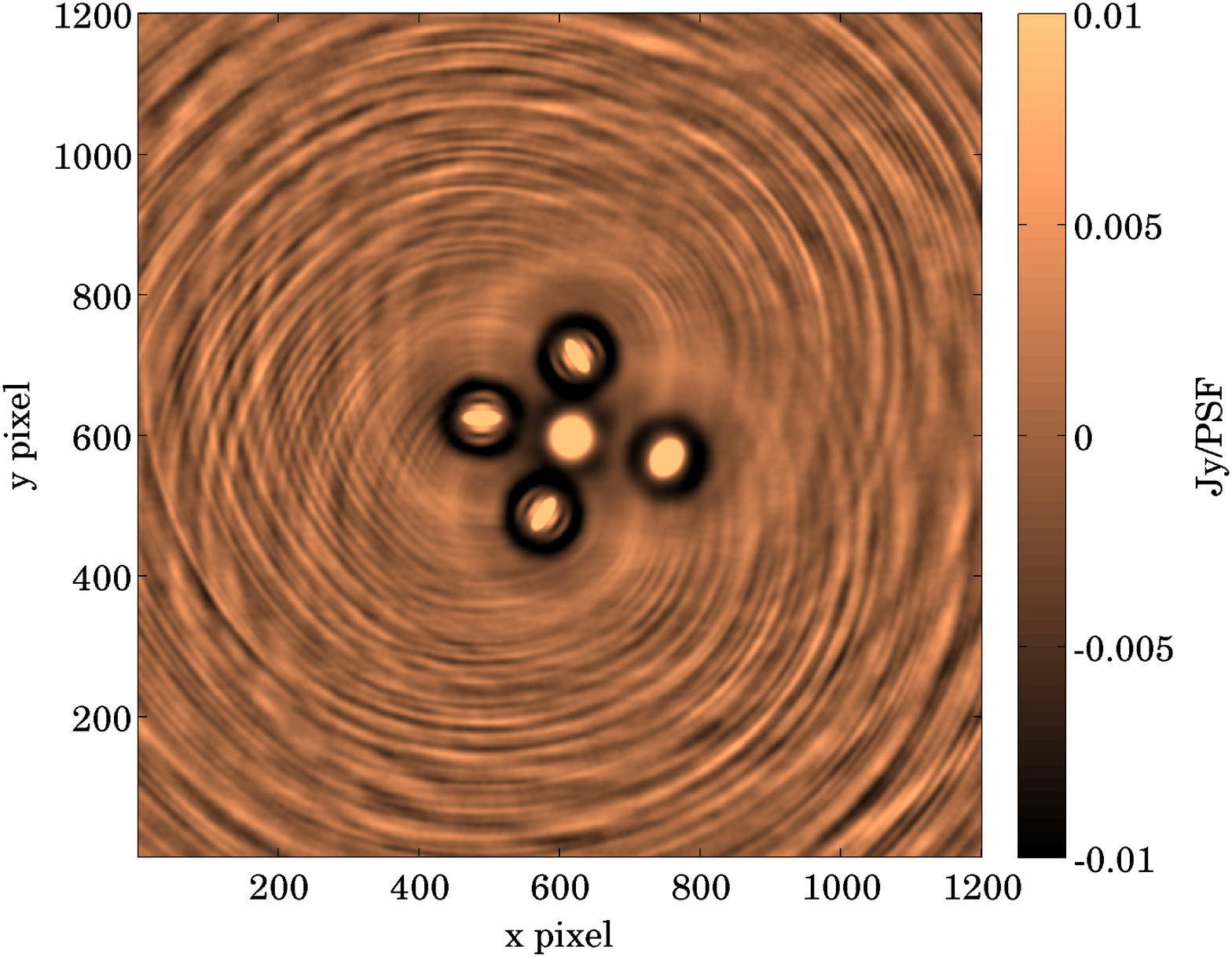,width=8.0cm}}
\vspace{0.5cm} \centerline{(c)}\smallskip
\end{minipage}
\begin{minipage}{0.48\linewidth}
\centering
 \centerline{\epsfig{figure=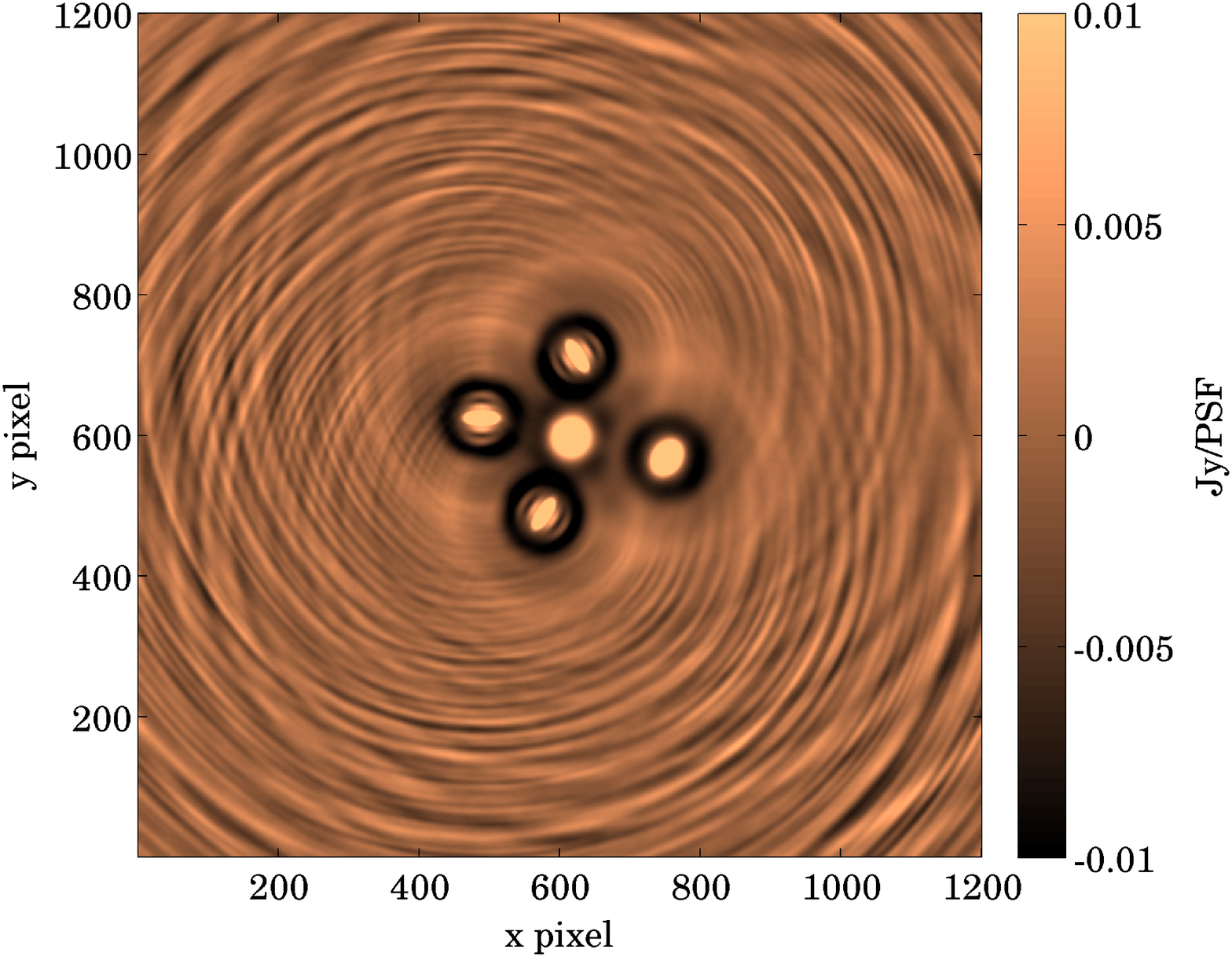,width=8.0cm}}
\vspace{0.5cm} \centerline{(d)}\smallskip
\end{minipage}
\begin{minipage}{0.48\linewidth}
\centering
 \centerline{\epsfig{figure=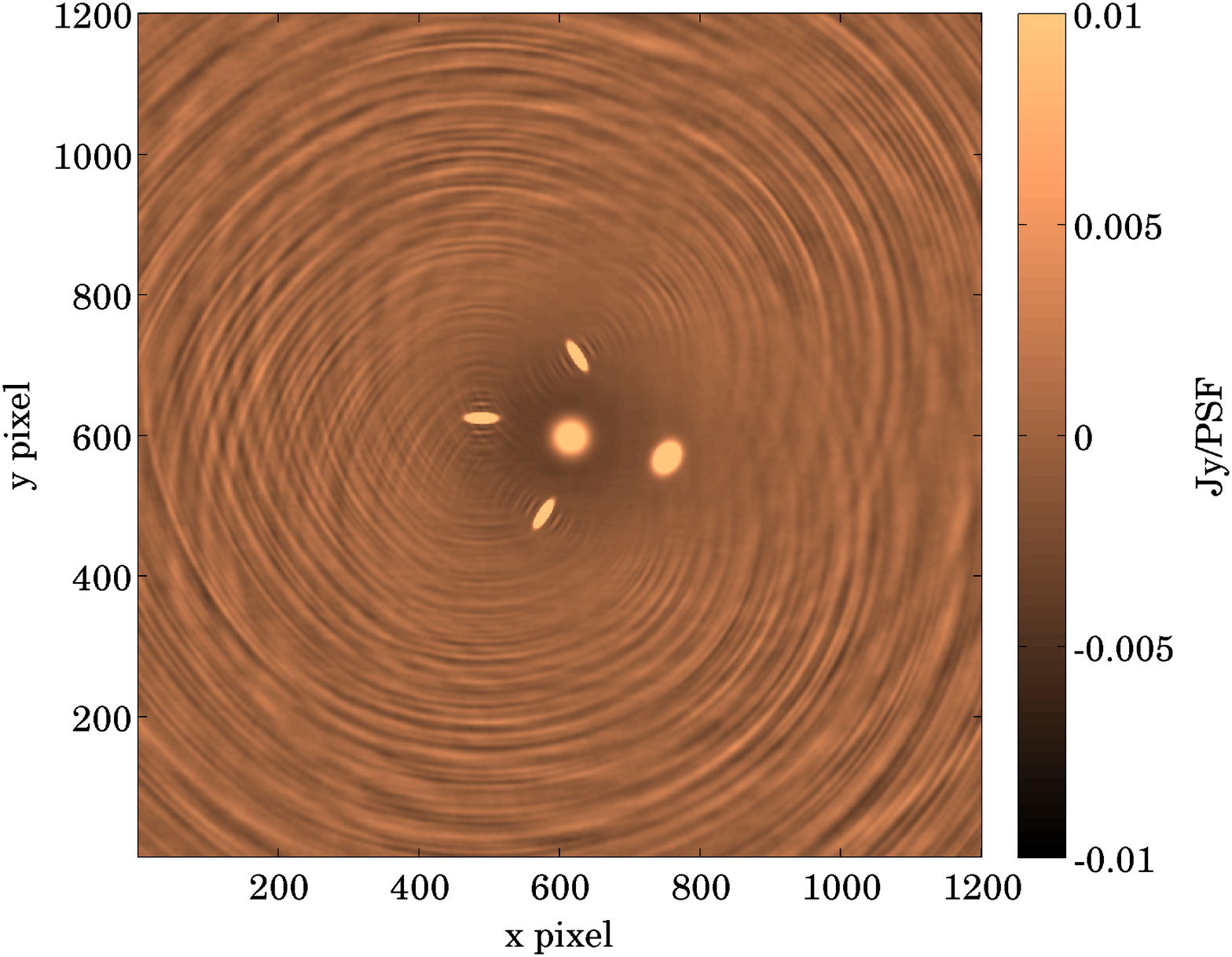,width=8.0cm}}
\vspace{0.5cm} \centerline{(e)}\smallskip
\end{minipage}
\end{center}
\caption{Simulated sky with 5 Gaussian sources at { 115} MHz, imaged using (a) Uniform weights (b) Natural weights (c) Gridding with uniform weights and tapering (d) Adaptive weights { (e) Briggs robust ($=0$) weights}. The FOV is 10$\times$10 square degrees, covered by 1200$\times$1200 pixels of size 30$^{\prime\prime}$.\label{sim_pm}}
\end{minipage}
\end{figure*}

We compare five weighting schemes in the simulations: (a) Uniform weights, (b) Natural weights, (c) Gridding the data using uniform weights and subsequently tapering the gridded data with aforementioned $G(u,v)$, (d) Adaptive weighting with aforementioned $G(u,v)$ with weights initialized to 1 and evaluating (\ref{Witer}) for 10 iterations, and { (e) Briggs (robust) weighting with robust parameter $=0$}. As an example, in Fig. \ref{psfcross} we show the PSF cross sections obtained by the different weighting schemes at { 115} MHz. In Fig. \ref{sim_pm}, we have shown images { (not deconvolved)} of the simulated sky with 5 Gaussian sources. The schemes (c) and (d) both give almost identical PSFs and images in Figs. \ref{psfcross} and \ref{sim_pm}, respectively. 

\begin{figure}
\begin{minipage}{0.98\linewidth}
\centering
 \centerline{\epsfig{figure=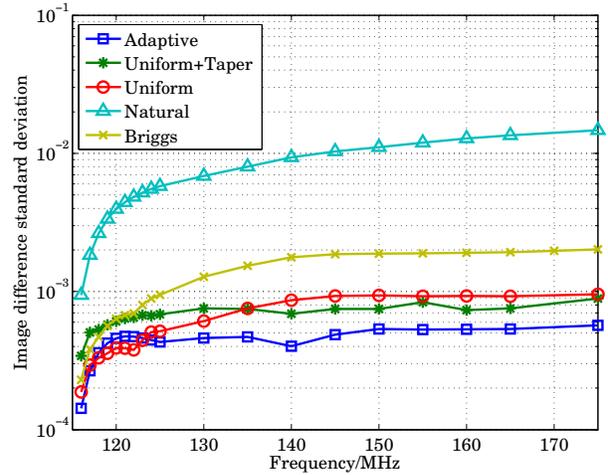,width=8.0cm}}
\caption{Standard deviation of the image difference compared to the image at 115 MHz. Adaptive weighing shows the least difference over most of the frequency range.\label{fidelity}}
\end{minipage}
\end{figure}

{ We simulate the same sky (no intrinsic variation) at different frequencies in the range 115 MHz to 175 MHz and make images using the five weighting schemes.  Note that we do not perform any image deconvolution. For each weighting scheme, we use the image at 115 MHz as our reference and find the standard deviation of the difference between this image and images made at other frequencies. Since there is no intrinsic variation in the sky model and moreover, since no image deconvolution is done, the only variation across the frequency range is due to the scaling of the uv sampling points. Therefore the standard deviation of the variation of images with frequency can be used as a measure of 'fidelity' and we have shown the results in Fig. \ref{fidelity}. Adaptive weighting shows the lowest difference except in a narrow band of frequencies where uniform weights give  a better result.}

\begin{figure}
\begin{minipage}{0.98\linewidth}
\centering
 \centerline{\epsfig{figure=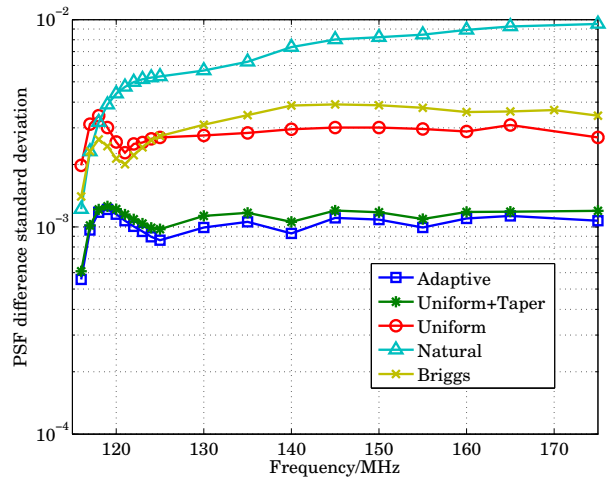,width=8.0cm}}
\caption{Standard deviation of the PSF difference compared with the PSF at 115 MHz, plotted against the frequency at which the PSF is calculated. Adaptive weighing shows the least difference in PSF.\label{psferror}}
\end{minipage}
\end{figure}

{
We also calculate the PSF at each frequency and find the standard deviation of the difference (compared with the PSF at 115 MHz) in the inner 400$\times$400 pixels.} We have shown the variation in the PSF difference with frequency in Fig. \ref{psferror}. Adaptive weighting shows the lowest variation in PSF as expected while natural weighting shows the largest variation.

\begin{figure}
\begin{minipage}{0.98\linewidth}
\centering
 \centerline{\epsfig{figure=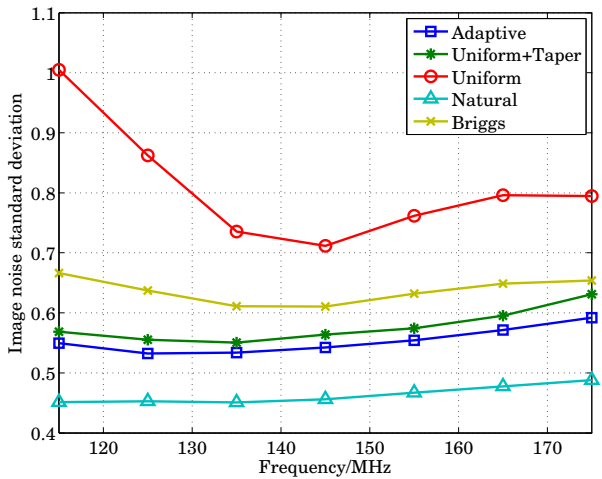,width=8.0cm}}
\caption{Image noise standard deviation for the five weighting schemes, plotted against frequency. Natural weights give lowest noise and Adaptive weights give the second lowest noise, { which is about 20\% higher}.\label{imgnoise}}
\end{minipage}
\end{figure}

In the next simulation, we simulate noise with equal variance at all baselines and calculate the noise standard deviation of images, that are made using the five different weighting schemes. We have shown the image noise variation with frequency in Fig. \ref{imgnoise}. While naturally weighted images give the lowest noise, adaptive weighting { increases the noise standard deviation by about 20\%.}

For the grand comparison, we multiply the mean values of the curves in Figs. \ref{psferror} and \ref{imgnoise} for the five weighting schemes. We normalize this such that adaptive weighting scheme has a value of 1. For schemes (a),(b) and (c) and (e), we get values of about 4.8, 4.2, 1.2 and 3.4, respectively. Therefore, { for the example considered,} the best weighting scheme that simultaneously minimizes PSF variation and maximizes sensitivity is the adaptive weighting scheme proposed in this paper. Gridding the data with uniform weights and thereafter tapering the data is about 20\% worse. We attribute this to the non-smooth variation of weights between different grid cells, as shown in Fig. \ref{Ex}. Considering the price that has to be paid in terms of computational cost to gain such a small improvement, one might question whether it is worth pursuing. However, for deep imaging experiments, even such a small improvement would prove critical. { Moreover, we emphasize that adaptive weighting alone is not sufficient to reach the noise limit in such deep imaging experiments. To minimize the effect of sources outside the field of view \citep{NCP2013}, directional calibration and source subtraction is essential.}

\section{Conclusions}\label{sec:conclusions}
In this paper, we have proposed an adaptive weighting scheme that can tune the PSF to match an externally given function. Simulations show the superiority of this weighting scheme in terms of minimizing PSF variation while maximizing sensitivity. Source code of an imager \citep{SYURSI2014} implementing this weighting scheme is available at http://exconimager.sf.net/. Future work in this topic would focus on improving computational cost and determining criteria for defining optimal functions { (and convolution kernels)} as input to the adaptive weighting scheme.

\section*{Acknowledgments}
We thank the reviewer, Melvyn Wright, for the careful review and insightful comments. We also thank the Editor and Assistant Editor for their comments. We also thank Ger de Bruyn for commenting on an earlier version of this paper.

\bibliographystyle{mn2e}
\bibliography{shapeletref}

\bsp
\label{lastpage}
\end{document}